\documentclass[useAMS,usenatbib,usegraphicx,dvips]{mn2e}
\title[3LAC {\it Fermi}-LAT Blazar Distributions]{A Determination of the Gamma-ray Flux and Photon Spectral Index Distributions of Blazars from the Fermi-LAT 3LAC}

\author[Singal]{J. Singal\thanks{E-mail: \texttt{jsingal@richmond.edu}}\\
Physics Department, University of Richmond \\ Gottwald Center for the Sciences \\ 28 Westhampton Way,  Richmond, VA 23173 \\
Accepted to MNRAS}
\begin{document}
\date{}
\pagerange{\pageref{firstpage}--\pageref{lastpage}} \pubyear{2015}
\maketitle
\label{firstpage}

\begin{abstract}
We present a determination of the distributions of gamma-ray photon flux -- the so called Log$N$-Log$S$ relation -- and photon spectral index for blazars, based on the third extragalactic source catalog of the {\it Fermi} Gamma-ray Space Telescope's Large Area Telescope, and considering the photon energy range from 100 MeV to 100 GeV.  The dataset consists of the 774 blazars in the so-called ``Clean'' sample detected with a greater than approximately seven sigma detection threshold and located above $\pm 20^{\circ}$ Galactic latitude.  We use non-parametric methods verified in previous works to reconstruct the intrinsic distributions from the observed ones which account for the data truncations introduced by observational bias and includes the effects of the possible correlation between the flux and photon index.  The intrinsic flux distribution can be represented by a broken power law with a high flux power-law index of -2.43$\pm$0.08 and a low flux power-law index of -1.87$\pm$0.10.  The intrinsic photon index distribution can be represented by a Gaussian with mean of 2.62$\pm$0.05 and width of 0.17$\pm$0.02.  We also report the intrinsic distributions for the sub-populations of BL Lac and FSRQ type blazars separately and these differ substantially.  We then estimate the contribution of FSRQs and BL Lacs to the diffuse extragalactic gamma-ray background radiation.  Under the simplistic assumption that the flux distributions probed in this analysis continue to arbitrary low flux, we calculate that the best fit contribution of FSRQs is 35\% and BL Lacs 17\% of the total gamma-ray output of the Universe in this energy range.

\end{abstract}


\section{Introduction} \label{intro}

In this work we investigate the distributions of gamma-ray photon flux and photon index for the largest flux-limited gamma-ray sample of blazars available at the moment, that of the third extragalactic source catalog of the {\it Fermi} Large Area Telescope (LAT) \citep[3LAC --][]{FermiAGN3}.  Most of the extragalactic objects observed by the Large LAT on the {\it Fermi} Gamma-ray Space Telescope are classified as blazars \citep[e.g.][]{Fermiyr1}, the active galactic nuclei (AGNs) in which the jet is pointing toward us \citep[e.g.][]{BK79}.  Characterizing the intensity and spectral characteristics of the gamma-ray emission seen in blazars is an essential for understanding of the physics of the accretion disk and black hole systems in AGNs \citep[e.g.][]{Dermer07}. Understanding the characteristics of blazars is also crucial for evaluating their contribution to the extragalactic gamma-ray background (EGB) radiation.  

The {\it Fermi}-LAT source catalogs are based on observations extending down to gamma-ray energies of 100 MeV.  Although the 3LAC reports the photon fluxes in the range from 1 to 100 GeV, the photon flux in the range from 100 MeV to 100 GeV is an important quantity for blazars because the most photons are at the lower energies of this range, as are the most photons of the EGB.  Therefore, we base this analysis on the full 100 MeV to 100 GeV photon flux, reconstructed as described in \S \ref{datasec}.  If the whole 100 MeV to 100 GeV information is used, the threshold flux for detection depends strongly on an object's gamma-ray spectrum.  This arises from the energy dependence of the point spread function \citep{Atwood09} and is such that harder spectra are detected at lower fluxes.  Thus, for determination of the flux distribution one needs to take into account both the flux and the photon spectral index, and that one then deals with a bi-variate distribution of fluxes and indexes, which is truncated because of the above mentioned observational bias.  Because the flux detection threshold depends on the photon index, the observed raw distributions do not provide the true Log$N$-Log$S$ counts or the true distribution of the photon index.  Thus a bias free determination of the distributions is more complicated than just counting sources.   Here we use non-parametric methods to determine the distributions directly from the data at hand.  Other methods such as monte carlo have been used, such as in \citet{Marco} -- hereafter MA, and, for example, with redshift information and suitable assumptions the distribution of photon index can be corrected for the bias towards hard spectrum sources as in \citet{V09}.

In a previous work (\citet{BP1} -- hereafter BP1) we demonstrated the techniques with real and simulated blazar flux and photon index data from the first year {\it Fermi}-LAT extragalactic source catalog \citep[1LAC --][]{FermiAGN}.  As explored by e.g. Petrosian (1992), when dealing with a bi-(or more generally multi)-variate distribution, one must determine the correlation (or statistical dependence) between the variables, which cannot be done by simple procedures when the data are truncated.  We use a method based on the techniques first developed by Efron and Petrosian \citep[EP;][]{EP92,EP99} which determine the intrinsic correlations (if any) between the variables and then the mono-variate distribution of each variable, accounting for the truncations of the data. These techniques have  been proven useful for application to many sources with varied characteristics and most recently to the gamma-ray flux and photon index of blazars in BP1 and to radio and optical luminosity in quasars in \citet{QP2}, where a more thorough discussion and references to earlier works are presented.  

In \citet{BP2} we used the spectroscopic redshift information provided for the complete sample of 1LAC Flat Spectrum Radio Quasar (FSRQ) type blazars to determine the gamma-ray luminosity function, its evolution, and the density evolution for FSRQs.  A very similar sample was used with different analysis techniques in \citet{Marco2} to obtain the gamma-ray luminosity function and evolution for FSRQs, and a 1LAC-based sample with a combination of measured redshifts and upper and lower limits was used to do the same with BL Lacs in \citet{Marco3}.  However, in the case of 3LAC blazars, the spectroscopic redshift information is incomplete, leading to the utility of this work recovering the intrinsic flux and photon index distributions using the maximal number of well characterized gamma-ray detected blazars of any type. 

In this paper we apply these methods to determine the correlation and the intrinsic distributions of flux and photon index of {\it Fermi}-LAT blazars.   In \S \ref{datasec} we discuss the data used from the 3LAC extragalactic catalog, and in \S \ref{dandr} we explain the techniques used and present the results.  In \S \ref{bgndradsec}  we describe how the results from such studies are important for understanding the origin of the extragalactic gamma-ray background (EGB) radiation. A discussion is presented in \S 5.

\section{Data}\label{datasec}

\begin{figure}
\includegraphics[width=3.5in]{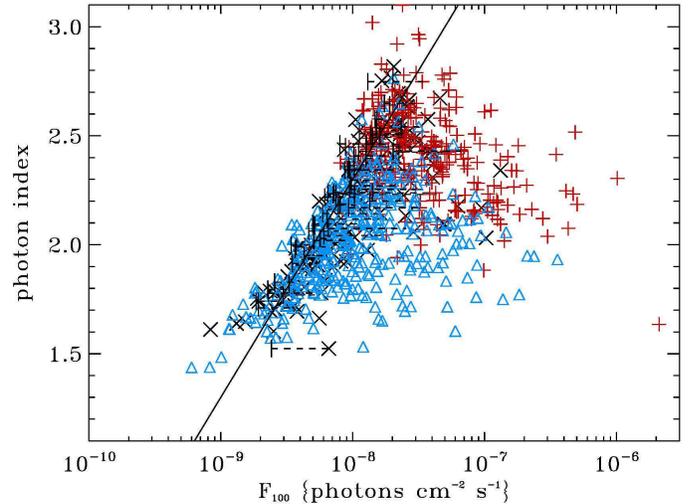}
\caption{ Flux and reported photon spectral index for the 774 {\it Fermi}-LAT blazars used in this analysis, those with test statistic $\geq$ 50 and $\vert b \vert \geq 20^\circ$.  BL Lac type blazars (n=352) are shown as blue triangles, FSRQs type blazars (n=286) are shown as red plus signs, and blazars of unidentified or ambiguous type (n=136) are represented by black x's.  It is seen that there is a selection bias against soft spectrum sources at fluxes below $\sim 6\times10^{-8}$ photons cm$^{-2}$ sec$^{-1}$.  We also show for a selection of sources (but only a few for clarity) the approximate limiting flux for that source -- that is the lowest flux it could have and still be sufficiently bright to be included in the sample given its location on the sky given the reported detection significance.  The location of the line used for the truncation boundary is also shown.   }
\label{lumsandsis}
\end{figure}

For this analysis we use the blazar sources reported in the {\it Fermi}-LAT 3LAC \citep{FermiAGN3} that are part of the ``Clean'' sample, have a detection test statistic $TS \geq 50$ and which lie at Galactic latitude $\vert b \vert \geq 20^{\circ}$.  This set of criteria, which has been adopted by the LAT team for certain analyses of the blazar population (e.g. MA), includes those sources that are fully calibrated, removes spurious sources, and maximizes the likelihood of sources being properly identified as to spectral type.  The test statistic is defined as $TS = -2 \, \times \, (\ln (L_0) - \ln (L_1))$, where $L_0$ and $L_1$ are the likelihoods of the background (null hypothesis) and the hypothesis being tested (e.g., source plus background).  The significance of a detection is approximately $n \times \sigma = \sqrt{TS}$ .  

These include 286 which are identified as FSRQ type, 352 which are identified as BL Lacertae (BL Lac) type, and 136 which have uncertain type.  This can be compared to the 1LAC with 352 $TS \geq 50$ and  $\vert b \vert \geq 20^{\circ}$ blazars, of which 161 are FSRQs, 163 are BL Lacs, and 28 which are identified as uncertain type.  Thus the dataset used in this analysis contains more than twice as many blazars as that used in the previous analyses of BP1 and MA.  There have also been some blazars that have been reclassified as a different type since the 1LAC, based on e.g. spectral data obtained by \citet{Shaw} and \citet{Shaw2} among other works.

The gamma-ray fluxesof the blazars in the range from 100 MeV to 100 GeV, designated here as $F_{100}$, and reported photon indexes are plotted in Figure \ref{lumsandsis}.   The 774 blazars range in $F_{100}$ from $6.01 \times 10^{-10}$ to $2.10 \times 10^{-6}$ photons cm$^{-2}$ sec$^{-1}$.  The photon index $\Gamma$ is defined such that for the photon spectral density at energy $E$, $n\!(E)dE \propto E^{-\Gamma}$ (or the $\nu F_{\nu} \propto \nu^{-\Gamma +2}$), and is obtained by the LAT collaboration by fitting a power-law to the observed spectra in the above energy interval.  In this dataset the reported photon index ranges from 1.438 to 3.100.  It should be noted that some blazars manifest a photon flux as a function of photon energy in the range from 100 MeV to 100 GeV that cannot be completely described by a single power law, showing curvature in the spectrum or other features.  However, the 3LAC does report a best-fit power law index for every blazar, and this analysis uses that power law.

We recover $F_{100}$ from the reported flux density ($K$), pivot energy ($E_p$), and photon index with

\begin{equation}
F_{100} = \int_{100 MeV}^{100 GeV} \, K \, {\left( E \over E_ p \right)}^\Gamma \, dE
\label{F100}
\end{equation}
$F_{100}$ is the blazar gamma-ray flux measure adopted in e.g. \citet{Marco} and is useful for exploring blazars and their relationship to the EGB as discussed in \S \ref{intro}.  We obtain the flux density and the pivot energy for 3LAC blazars from the corresponding general {\it Fermi} source catalog \citep{Fermi3}.  The bias mentioned above is clearly evident; there is a strong selection against soft spectrum sources at fluxes below $F_{100} \sim 6\times 10^{-8}$ photons cm$^{-2}$ sec$^{-1}$. 

Every blazar has an associated $TS$ as discussed above, with the background flux being a function of position on the sky \citep{FermiAGN}.  In Figure \ref{lumsandsis} we also show the approximate limiting flux of some (only some to avoid visual confusion) of the blazars, which is an estimate of the lowest flux it could have to be included in the sample, given by $F_{\rm lim}=F_{100} / \sqrt{TS/50}$.  This $F_{\rm lim}$ will thus have some dependence on the object's photon index and position on the sky.  However, as discussed in BP1, since there is some uncertainty in the detection threshold values of fluxes and indexes, the limiting flux as determined in this way is not the optimal estimate.  We therefore use a more conservative truncation as shown by the straight line in Figure \ref{lumsandsis}.  As verified with the simulated data discussed in BP1, moving the truncation line to the right and down eliminates more sources in the edges of the sample, and does not change the results but increases the uncertainty, while moving it to the left changes the results.  We choose the truncation line boundary that is at the edge of this region where the results are not changed.  This way we lose some data points but make certain that we are dealing with a complete sample with a well defined truncation.

\section{Determinations of Distributions}\label{dandr}

\subsection{Correlations}\label{corrs}

\begin{figure}
\includegraphics[width=3.5in]{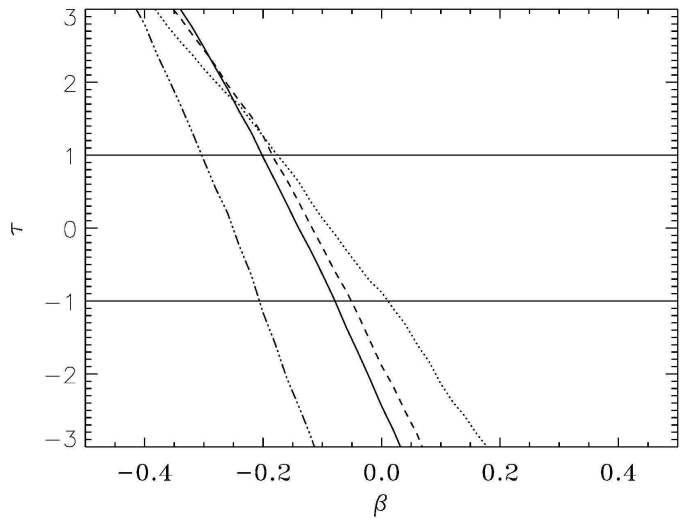}
\caption{Test statistic $\tau$ versus intrinsic correlation factor $\beta$ for a photon index and flux correlation of the form given in Equation \ref{fdef}, for the 3LAC blazars (solid curve), the subset of BL Lac type blazars (dashed curve), and the subset of FSRQs type blazars (dash-dot curve).  The 1$\sigma$ range of best fit values for $\beta$ are where $\vert \tau \vert \leq 1$.  For comparison, the dotted curve shows the correlation factor for just those sources above $6\, \times \,10^{-8}$  photons cm$^{-2}$ sec$^{-1}$, where the data truncation in the $F_{100}$,$\Gamma$ plane is not as relevant. }
\label{corrfig}
\end{figure}

As discussed in \S \ref{intro}, with a bi-variate truncated dataset one must determine whether the variables are independent before considering distributions.  If in this case $F_{100}$ and $\Gamma$ are independent, the combined distribution $G\!(F_{100},\Gamma)$ can be separated into two independent distributions $\psi\!(F_{100})$ and $h\!(\Gamma)$.  However, they may not be independent even though flux clearly depends strongly on redshift while photon index may not, and in fact does not in the case of FSRQs \citep{BP2}.  Correlations between flux and photon index are observed for example in gamma-ray bursts \citep{Yonetoku04,Lloyd00}.  There may also be an observed correlation induced by the selection biases, as there certainly is in this case as can be seen in Figure \ref{lumsandsis}, which should be removed in order to obtain bias free distributions of the variables. Thus, the first task is to establish whether the variables are independent. 

While the flux and photon index exhibit a strong correlation in the raw (and heavily biased) data, determining the intrinsic correlation when the data are truncated requires statistical methods to account for the missing data.  In BP1 we discuss in detail the methods we apply to the bi-variate {\it Fermi}-LAT observational blazar data.  We use a technique first explored by Efron and Petrosian \citep{EP92,EP99} to determine whether the two variables are correlated. This method utilizes a modified version of the Kendall Tau test with the test statistic

\begin{equation}
\tau = {{\sum_{j}{(\mathcal{R}_j-\mathcal{E}_j)}} \over {\sqrt{\sum_j{\mathcal{V}_j}}}}
\label{tauen}
\end{equation}
to quantify the independence of two variables in a dataset, say ($x_j,y_j$) for  $j=1, \dots, n$.  Here $\mathcal{R}_j$ is the dependent variable ($y$) rank of the data point $j$ in a set associated with it.  For untruncated data (i.e. data truncated parallel to the axes) the set associated with point $j$ includes all of the points with a lower (or higher) independent variable value ($x_k < x_j$).  If the data is truncated the unbiased set is then the {\it associated set} consisting only of those points of lower(or higher) independent variable ($x$) value that would have been observed if they were at the $x$ value of point $j$ given the truncation.  

If ($x_j,y_j$) are uncorrelated then the ranks of all of the points $\mathcal{R}_j$ in the dependent variable within their associated set should be distributed uniformly between 0 and the number of points $n$, with the rank uncorrelated with their independent variable value, with the expectation value $\mathcal{E}_j=(1/2)(n+1)$ and variance  and $\mathcal{V}_j=(1/12)(n^{2}-1)$, where $n$ is the number of objects in object $j$'s associated set.  Then the points' contributions to $\tau$ will tend to sum to zero.   On the other hand, if the indepenent and dependent variables are correlated, then the rank of a point in the dependent variable will be correlated with its  independent variable, and because the set to be ranked against consists of points with a lower independent variable value, the contributions to $\tau$ will not sum to zero.

Independence of the variables is rejected at the $m \, \sigma$ level if $\vert \, \tau \, \vert > m$, and this can be considered the same standard deviation as would be calculated from another method such as least-squares fitting, as discussed in \citet{EP99}.   If the variables are not independent, to find the best fit correlation the $y$ data are then adjusted by defining some new dependent variable $y'_j=F(y_j,x_j)$  and the rank test is repeated, with different values of parameters of the function $F$.

 To determine the intrinsic correlation between $F_{100}$ and $\Gamma$ we use a function which is a simple coordinate rotation, defining a new variable we call the ``correlation reduced photon index'' as

\begin{equation}
\Gamma_{\rm cr}=\Gamma - \beta\, \times\, \log\left({ {F_{100}} \over {F_0} }\right).  
\label{fdef}
\end{equation}
Then we determine the value of the parameter $\beta$ empirically that makes $F_{100}$ and $\Gamma_{\rm cr}$ independent.  This is the best fit value of the correlation between $F_{100}$ and $\Gamma$ for this functional form.  The distributions of $F_{100}$ and $\Gamma_{\rm cr}$ are indeed separable:

\begin{equation}
G\!(F_{100},\Gamma) = \psi\!(F_{100}) \, \times \, \hat h\!(\Gamma_{\rm cr}). 
\end{equation}

Once the monovariate distributions of $F_{100}$ and $\Gamma_{\rm cr}$ are determined then the true distribution of $\Gamma$ can be recovered by an integration over $F_{100}$ as:

\begin{eqnarray}
h\!(\Gamma) =  
\nonumber \\
\int_{F_{100}}  \psi\!(F_{100}) \, \hat h\left(\Gamma-\beta \times \log\left({{F_{100}} \over F_0 } \right) \right) \,  d\,F_{100}.
\label{inteq}
\end{eqnarray}
Here $F_0$ is some fiducial flux we chose to be $F_0=6\times 10^{-8}$ photons s$^{-1}$ cm$^{-2}$ sr$^{-1}$ which is approximately where the flux distribution breaks (see below), although its actual value is not important -- it is simply utilized to avoid taking the logarithm of a dimensionless number.  The data described in \S \ref{datasec} are truncated in the $F_{100}$-$\Gamma$ plane, due to the bias against low flux, soft spectrum sources.   As discussed in \S \ref{datasec} we can use a curve approximating the truncation, $\Gamma_{\rm lim}=g\!(\log F_{100})$, which defines the associated set for each point as those objects whose photon index is less than the limiting photon index of the object in question given the truncation curve and its specific value of $F_{100}$. 

As discussed in BP1 we have tested this procedure using a simulated dataset from the {\it Fermi}-LAT collaboration that resemble the real observations and are subject to a truncation similar to the actual observational data, but with known input distributions of uncorrelated photon index and flux.  Note that when defining new variables the truncation curve as a function of flux should also be transformed  by same parameter $\beta$; 
\begin{equation}
\Gamma_{\rm cr, lim}=\Gamma_{\rm lim} - \beta\, \times\, Log\left({ {F_{100}} \over {F_{100 - min}}}\right).    
\end{equation}

Figure \ref{corrfig} shows the value of the test statistic $\tau$ as a function of the correlation parameter $\beta$ for a all  blazars and the subsets including only BL Lacs and FSRQs in the sample.  Table 1 shows the best fit values and 1$\sigma$ ranges of the correlation parameter $\beta$ for these cases.  We note that the correlation is generally small, for example for all {\it Fermi} blazars the best fit value is $\beta$=-0.13 $\pm$ 0.06, but significant.  This is not surprising, as there is some evidence of an inverse correlation between blazar luminosity and photon index \citep{Marco3}.  In \citet{BP2} we found no evidence of significant evolution of photon index with redshift for the case of FSRQs, so the result of a possible negative correlation seen here between photon index and flux would indicate that the (small) intrinsic correlation is actually indeed between photon index and luminosity.

\subsection{Distributions}\label{dist}

With the correlation removed the independent distributions $\psi\!(F_{100})$ and $\hat h\!(\Gamma_{\rm cr})$ can be determined using a method outlined in \citet{P92} based on techniques first devised by \citet{L-B71}.  These methods give the cumulative distributions by summing the contribution from each point without binning the data, and as discussed in \citet{L-B71}, are equivalent to the maximum likelihood distributions.

\subsubsection{flux distributions}\label{fdists}

\begin{figure}
\includegraphics[width=3.5in]{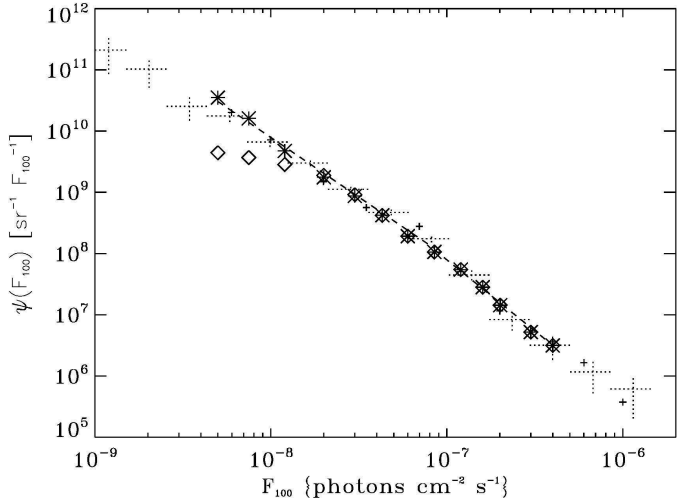}
\caption{ Observed (diamonds) and reconstructed intrinsic (stars) differential distribution of gamma-ray flux $\psi\!(F_{100})$ for the 3LAC {\it Fermi}-LAT blazars.  The vertical error bars on the stars represent the 1$\sigma$ range of the correlation parameter $\beta$, the dominant source of uncertainty, and are generally smaller than the points.  The intrinsic distribution is a power law with a break at $F_{\rm br}\simeq 8\, \times \,10^{-8}$  photons cm$^{-2}$ sec$^{-1}$.  The best fit slopes for the intrinsic distribution are -2.43$\pm$0.08 above the break and -1.87$\pm$0.10 below, and the best fit intrinsic distribution is plotted as the dotted line.  We also plot the best-fitting $\psi\!(F_{100})$ as determined in BP1 (small crosses) and MA with error bars (dotted lines). The best fit value for $\psi\!(F_{break})$ is 1.66 $\times 10^8$ sr$^{-1} F_{100}^{-1}$. The distribution parameters for the BL Lac and FSRQ sub-populations separately are presented in Table I. }
\label{fluxdistfig}
\end{figure}

For the flux, the cumulative distribution

\begin{equation}
\Phi\!(F_{100}) \equiv \int_{F_{100}}^\infty {\psi\!(F_{100}') \, dF_{100}'}
\end{equation}
is obtained with

\begin{equation}
\Phi\!(F_{100}) = \prod_{j} \left({1 + {1 \over N\!(j)}} \right)
\label{phieq}
\end{equation}
where $j$ runs over all objects with fluxes $F_{100,j}\geq F_{100}$, and $N(j)$ is the number of objects $i$ in the {\it associated set} of object $j$ that have a value of $F_{100,i} \geq F_{100,j }$.  The associated set for object $j$ in this case is all of those objects that have $\Gamma_{\rm cr} \leq \Gamma_{\rm cr, lim}(F_{100,j})$, determined from the truncation curve described above.   Equation \ref{phieq} represents the established Lynden-Bell method which we modify by including only those objects within object $j$'s associated set for the calculation of $N(j)$.  The use of only the associated set for each object removes the biases introduced by the truncation in this calculation.  

We determine the differential flux distribution 
\begin{equation}
\psi\!(F_{100}) = - {d \Phi\!(F_{100}) \over dF_{100}}
\label{psieqn}
\end{equation}
by fitting piecewise polynomial functions via least-squares fitting to $\Phi\!(F_{100})$ and calculating the derivatives.  Figure \ref{fluxdistfig} shows the calculated intrinsic differential distribution $\psi\!(F_{100})$, along with those obtained from the raw observed data without correcting for the bias.  A direct comparison to the results from BP1 and MA is presented there as well.   For consistency with earlier works such as MA we fit the differential counts as a broken power law which describes the data well and takes the form

\begin{eqnarray}\label{bpowlaw}
\psi\!(F_{100}) = \,\,\,\,\, \psi\!(F_{break}) \, \left( {  {{F_{100}} \over {F_{break}}} } \right)^{m_{above}}  \,\,\,\,\,{\rm for} \,\,\,\,\,F_{100} \geq F_{break} \\
\nonumber  \psi\!(F_{break}) \, \left( {  {{F_{100}} \over {F_{break}}} } \right)^{m_{below}} \,\,\,\,\,{\rm for} \,\,\,\,\,F_{100}< F_{break}.
\end{eqnarray}
$m_{above}$ and $m_{below}$ are the power law slopes above and below the break, respectively, and are obtained from a least-squares fitting of $\psi\!(F_{100})$, as is the value of $F_{break}$.  At values of $F_{100}$ above $F_{NT} \equiv 6\times 10^{-8}$  photons cm$^{-2}$ sec$^{-1}$ the truncation imposed by the boundary line shown in Figure \ref{lumsandsis} is not significant and we can obtain the normalization by scaling the cumulative distribution $\Phi\!(F_{100})$ such that  

\begin{equation}
\Phi\!(F_{NT}) = { {N \pm \sqrt{N}} \over {8.26 \, {\rm sr} } } , 
\label{normeqn}
\end{equation}
where $N$ is the number of objects above $F_{NT}$ in the dataset in question and is equal to 89 for all blazars, 20 for BL Lacs, and 65 for FSRQs. The $\sqrt{N}$ uncertainty arises because of Poisson noise for the brightest sources, and 8.26 sr is the total sky coverage considered, which is $\vert b \vert \geq 20^{\circ}$ as discussed in \S \ref{datasec}.  This normalization then allows the properly normalized value of $\psi\!(F_{break})$ to be calculated.  The best fit value for $\psi\!(F_{break})$, corresponding to the best-fitting value of $m_{above}$ is then 1.2 $\times 10^8$ sr$^{-1} F_{100}^{-1}$. 

\subsubsection{Photon index distributions}\label{pdist}

\begin{figure}
\includegraphics[width=3.5in]{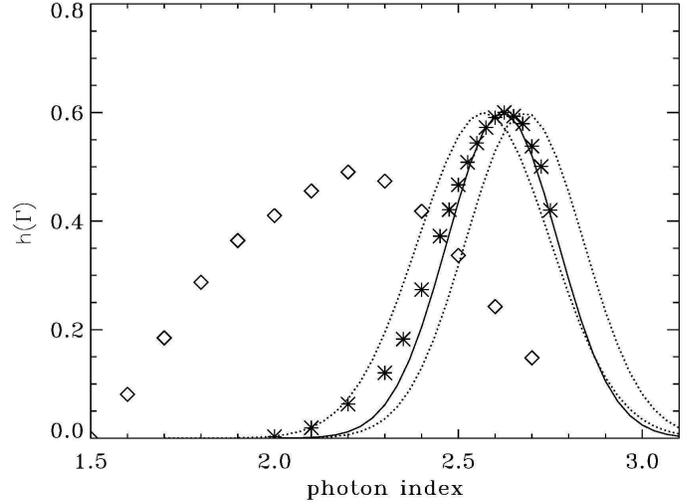}
\caption{Observed (diamonds) and reconstructed intrinsic (stars) distribution of photon index $h\!(\Gamma)$ for the 3LAC {\it Fermi}-LAT blazars used in this analysis. The intrinsic distribution is calculated from the flux distribution and the correlation reduced photon index distribution by equation \ref{inteq}.  The stars represent the intrinsic distribution calculated with the best fit value of the correlation parameter $\beta$ and the solid curve is the best fit Gaussian function to these values, while the dotted curves represent the best fit Gaussian functions to the extremal intrinsic distributions allowed by the 1$\sigma$ range of $\beta$.  The intrinsic distribution can be represented by a Gaussian with a mean of 2.62$\pm$0.05 and 1$\sigma$ width of 0.17$\pm$0.02, while the raw observed distribution can be represented by a Gaussian with a mean of 2.19$\pm$0.01 and 1$\sigma$ width of 0.34$\pm$0.01.  The normalization of $h\!(\Gamma)$ is arbitrary.}
\label{adistfig}
\end{figure}

To determine the cumulative distribution of the correlation reduced photon index

\begin{equation}
\hat P\!(\Gamma_{\rm cr}) \equiv \int_0^{\Gamma_{\rm cr}} {\hat h\!(\Gamma_{\rm cr}') \, d\,\Gamma_{\rm cr}'}
\end{equation}
we use

\begin{equation}
\hat P\!(\Gamma_{\rm cr}) = \prod_{k} \left({1 + {1 \over M\!(k)}} \right)
\end{equation}
in the modified method of \citet{L-B71} as above, which can be differentiated to give the differential distribution

\begin{equation}
\hat h\!(\Gamma_{\rm cr}) = {d \hat P\!(\Gamma_{\rm cr}) \over d\,\Gamma_{\rm cr}}
\end{equation}
In this case, $k$ runs over all objects with a value of $\Gamma_{\rm cr, k}\leq \Gamma_{\rm cr}$, and $M(k)$ is the number of objects $i$ with $\Gamma_{cr, i} \leq \Gamma_{cr, k}$ in the {\it associated set} of object $k$.  Here the associated set for object $k$ is those objects with $F_{100} \geq F_{\rm lim,k}$ obtained from the truncation line at $\Gamma_{\rm cr,k}$.  We note again that the cutoff curve as a function of flux is scaled by $\beta$ in the same manner of equation \ref{fdef}. 

${\hat h}\!(\Gamma_{\rm cr})$ can be used via equation \ref{inteq} to obtain the intrinsic distribution of the photon index itself, $h\!(\Gamma)$. The results are shown in Figure \ref{adistfig} along with the raw observed distribution for comparison.  Because the mean of intrinsic distribution of photon index is sensitive to the value of the correlation parameter $\beta$, we include the full range of intrinsic distributions resulting from the 1$\sigma$ range of $\beta$. A Gaussian form provides a good description of the intrinsic distribution of the index.  

We have carried out the same procedures to obtain the distributions of the BL Lac and FSRQ subsets of the data.  Table 1 summarizes the best fit parameters for the intrinsic flux and photon index distributions, for the sample considered as a whole, and for the BL Lac and FSRQs sub-populations separately.  The errors reported include statistical uncertainties in the fits and the deviations resulting from the 1$\sigma$ range of the correlation parameter $\beta$, with the later being much larger in magnitude.  A higher value of $\beta$ (i.e. more positive correlation between flux and photon index absolute value) moves the mean of the photon index distribution down to a lower absolute value of the photon index and makes the faint end source counts slope less steep (less negative $m_{below}$), while a lower value of $\beta$ has the opposite effect.

\section{Total Output from Blazar populations and the Extragalactic Gamma-ray Background}
\label{bgndradsec}

\begin{figure}
\includegraphics[width=3.5in]{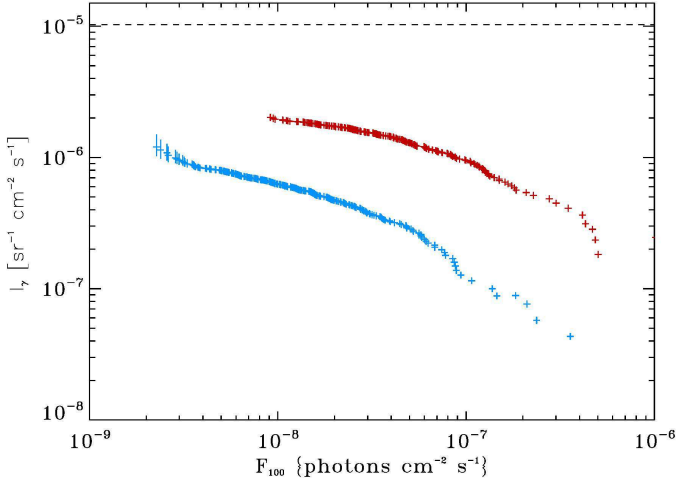}
\caption{ Estimate of the cumulative number of photons between 0.1 and 100 GeV above a given $F_{100}$,  $\mathcal{I}_{\rm \gamma - pop}(>F_{100})$ from equation \ref{CGB2}, from FSRQs (red upper points) and BL Lacs (blue lower points), shown with error bars resulting from the 1$\sigma$ range of the correlation parameter $\beta$.  The error bars are generally smaller than the plotting symbols used, except for the leftmost points.  The dashed horizontal line shows the EGB, ie the total extragalactic gamma-ray output ($\mathcal{I}_{EGB}$) as defined here.  An estimated total contribution of a population to $\mathcal{I}_{EGB}$ can be obtained by integrating equation \ref{CGB2} to zero flux, as discussed in \S \ref{bgndradsec}.  }
\label{bgndcontfig}
\end{figure}

One can integrate the flux distribution to calculate the total flux from a population and its proportional contribution to a photon background.  In this case, we are interested in the EGB, defined here as the total extragalactic gamma-ray photon output.\footnote{This definition avoids the problem that individual instruments resolve a different fraction of sources, leading to different estimates for the fraction of the total extragalactic photon output that is unresolved.}  In BP1 we focused on using the flux distribution to estimate the contribution from blazars as a whole.  Here with a much larger sample we will use the flux distributions to estimate the FSRQ and BL Lac contributions separately.  The total output in gamma-ray photons from blazar sources with fluxes greater than a given flux $F_{100}$, in terms of photons s$^{-1}$ cm$^{-2}$ sr$^{-1}$ between 0.1 and 100 GeV, is 

\begin{equation}
\mathcal{I}_\gamma(>F_{100})=\int_{F_{100}}^{\infty} \, F'_{100} \, \psi\!(F'_{100}) \, dF'_{100}.
\label{CGB1}
\end{equation}
Integrating by parts the contribution to the EGB can be related directly to the cumulative distribution $\Phi\!(F_{100})$ which is the primary output of our procedure
\begin{equation}
\mathcal{I}_\gamma(>F_{100})=F_{100} \, \Phi\!(F_{100})+\int_{F_{100}}^{\infty} \Phi\!(F'_{100}) \, dF'_{100}.
\label{CGB2}
\end{equation}
The advantage of using the latter equation is that it can give a step-by-step cumulative total contribution to the background instead of using analytic fits to the differential or cumulative distributions obtained from binning the data.  Figure \ref{bgndcontfig} shows $\mathcal{I}_\gamma(>F_{100})$ for FSRQs and BL Lacs {\it at fluxes probed by this analysis}.  Note that this includes the contribution from both detected blazars and some of those undetected owing to the truncation in the $F_{100}, \Gamma$ plane but still probed by the analysis.  Therefore, this calculated contribution can be more than the total contribution of blazars resolved by {\it Fermi}-LAT in the 3LAC.

In order to estimate the contribution of objects at fluxes {\it not} probed by the {\it Fermi}-LAT to the total EGB one can extrapolate the flux distribution we have obtained to lower fluxes.  This extrapolation is more uncertain. We fit a power law to the faint end of the $\Phi\!(F_{100})$ distributions so that we can extend the integration of equation \ref{CGB2} to lower fluxes.  Integrating to zero flux we find that FSRQs would produce a photon output of $\mathcal{I}_{\gamma}(>F_{100}=0)$=3.6 (+3.4/-0.9) $\times 10^{-6}$ photons s$^{-1}$ cm$^{-2}$ sr$^{-1}$ and BL Lacs would produce a photon output of $\mathcal{I}_{\gamma}(>F_{100}=0)$=1.8 (+4.5/-0.5) $\times 10^{-6}$ photons s$^{-1}$ cm$^{-2}$ sr$^{-1}$ .  These relatively large ranges are due to the uncertainty in the faint end cumulative source counts slope, ultimately owing to the range of the correlation parameter $\beta$, where the best fit value reported is for the middle of the 1$\sigma$ faint end slope of $\Phi\!(F_{100})$.  

This is to be compared with the total observed EGB, consisting of unresolved emission plus resolved sources, of $\mathcal{I}_{EGB}=10.4 \times 10^{-6}$ photons s$^{-1}$ cm$^{-2}$ sr$^{-1}$ reported by {\it Fermi} \citep{Fermibgnd3}.  It should be noted that this represents a change from the previous {\it Fermi} estimate of $\mathcal{I}_{EGB}=14.4\pm1.9 \times 10^{-6}$ photons s$^{-1}$ cm$^{-2}$ sr$^{-1}$ \citep{Fermibgnd}.  {\it If} the populations continue to have the fitted power law distribution to zero flux then FSRQs would produce 35(+35/-9)\% of the observed EGB and BL Lacs would produce 17(+44/-12)\%.  

These results are roughly consistent with a number of other findings.  In \citet{BP2} we consider redshift evolution effects with a smaller sample with complete available spectroscopic redshifts and find that for FSRQs the contribution to the EGB is 22(+10/-4)\%, with the result as reported here scaled by a factor of 1.38 for the ratio of the higher previously reported {\it Fermi} EGB intensity to the new recently reported lower intensity, while \citet{Marco2} report 30.1(+2.5/-1.7)\% for FSRQs with the result as reported here likewise scaled.  For BL Lacs, \citet{Marco3} estimate that they contribute between 14\%-20\% of the EGB with the same scaling\footnote{The original reported numbers are 16(+10/-4)\% for FSRQs from BP1, 21.7(+2.5/-1.7)\% for FSRQs from \citet{Marco2}, and 10\%-15\% for BL Lacs from \citet{Marco3}.}.  However, comparing the results obtained here from a large sample but with only flux information to those considering smaller samples with redshift evolution information can be enlightening, as discussed in \S \ref{discsec}.

We note that blazars should not be present to arbitrarily low flux so the integration here to zero flux is an overestimate in particular for the high ranges of the above estimates which are most sensitive to the low flux integration limit.  The best fit values obtained here do not favor blazars being the sole significant contributor to the EGB.  Several authors \citep[e.g.][]{SV11,A10} have suggested that blazars could be the primary source of the EGB. The spectral index of the EGB of $\sim$2.4 \citep{Fermibgnd} is consistent with the mean photon index of the blazars as determined in BP1 and in MA (although less so with the mean photon index determined here).  In a similar vein, \citet{VP11} have shown that the spectrum of the EGB is consistent with a blazar origin.  Other possible source populations include starforming galaxies, which have been recognized as a possible major contributor to the EGB by e.g. \citet{SV11}, \citet{Fields10}, and \citet{L11}, although this has been disputed by \citet{M11}, radio galaxies \citep[e.g][]{I11}, and other non-blazar AGN \citep[e.g.][]{IT09,IT11}.  Additional arguments against blazars being the sole significant contributor to the EGB have been made based on the observed EGB anosotropy in \citet{Cuoco12} and \citet{HA12}.

\section{Discussion}\label{discsec}

We have applied a method to calculate the intrinsic distributions in flux and photon index of blazars directly from the observed {\it Fermi}-LAT 3LAC catalog without reliance on simulations.  This method accounts for the pronounced data truncation introduced by the selection biases inherent in the observations when the full 100 MeV to 100 GeV observational range is used, and addresses the possible correlation between the variables. The accuracy of the methods used here when applied to {\it Fermi}-LAT blazar flux and photon index data were demonstrated in the Appendix of BP1 using a simulated dataset with known distributions.  A summary of the best fit correlations between photon index and flux, and the best fit parameters describing the inherent distributions of flux and photon index, are presented in Table 1 along with the values obtained in BP1 with the smaller 1LAC dataset and by MA.  We have obtained the intrinsic distributions considering the major data truncation arising from {\it Fermi}-LAT observations.  More subtle issues affecting the distributions we have derived, especially the photon index distribution, may arise due to the finite bandwidth of the {\it Fermi}-LAT and lack of complete knowledge of the objects' spectra over a large energy range and deviations from simple power laws.  However the {\it Fermi}-LAT bandwidth is sufficiently large that the contribution of sources which peak outside of this range to the source counts and the EGB in this energy range will be small.  

A limitation of these results is the use of the reported best-fit power-law photon index for each blazar, even though some, or perhaps most, sources actually deviate from a strict power law in the relation between photon flux and photon energy.  According to \citet{FermiAGN3} a total of 91 FSRQs, 32 BL Lacs, and 8 blazars of unknown type in the 3LAC show significant curvature in their spectra.  As shown in that paper it is overwhelmingly the highest $TS$ (and therefore generally higher flux) blazars which have been shown to have significant curvature away from a true power law in this relationship.  Spectral curvature could be a property of the higher flux blazars, or possibly more likely it is a general feature of blazars which has been most readily observable in those blazars with the highest TS.  If the former is the case, then the effect on the truncation and issues related to the truncation would be minimal no matter how the truncation is dealt with.  However, even in the case of the later, since this analysis  relies on an empirical curve to approximate the truncation in the reported F$_{100}$,$\Gamma$ plane as discussed in \S \ref{corrs}, the analysis itself should not be particularly sensitive to deviations from a strict power law.  However, to the extent that power laws are not representative of the true spectral properties of bright blazars in the energy range from 100 MeV to 100 GeV, that will be the case in these results and the distributions of photon index must be understood to be distributions of the {\it best fit} power law photon index.  As stated in \citet{Spectra}, ``Although some spectra display significant curvatures, the photon index obtained by fitting single power-paw models over the whole LAT energy range provides a convenient means to study the spectral hardness.''

In \S 3.2 of BP1 we consider the uncertainties and errors present in this analysis resulting from individual measurement uncertainties, blazar variability, and source confusion.  We conclude that the errors and uncertainties introduced from each of these effects are small compared to the dominant source of uncertainty in this analysis, which is that arising from the range of the correlation parameter $\beta$ which propagates through every other determination.  As we argue in BP1, to the extent that faint blazars are more likely to be observed by instruments such as the {\it Fermi}-LAT if they are in the flaring state rather than the quiescent state \citep[e.g.][]{SS96}, then the observed blazars should have a different mean photon index than the EGB, were the EGB to be made primarily from quiescent state blazars, under the assumption that blazars in the flaring state have a different spectrum than in the quiescent state.  As the reconstructed mean photon index here (and elsewhere) of the {\it Fermi}-LAT observed population is close to (although not exactly) that of the EGB, and there is only a weak relation and correlation between flux and photon index, this would imply that at least one of the following must be the case: a) there is not a significant bias in the {\it Fermi}-LAT toward detecting blazars in the flaring state, b) quiescent blazars do not form the bulk of the EGB, or c) flaring and quiescent blazars have, en masse, roughly the same photon index distributions. 

We find that the photon index and flux show a slight negative correlation.  This correlation is greater than 1$\sigma$ significance for all blazars, and is quite significant in the case of the FSRQ sub-population.  This correlation likely results from an underlying correlation between photon index and luminosity, as discussed in \S \ref{corrs}.  Viewing the flux and photon index distributions, the comparison of the intrinsic and raw observed distributions show clearly the substantial effects of the observational bias, as can be seen in Figures \ref{fluxdistfig} and \ref{adistfig}.  The intrinsic differential counts can be fitted adequately by a broken power law and the photon index appears to have an intrinsic Gaussian distribution. 

We also find that in general the values reported here are consistent with those reported in BP1 and MA for the power law slopes of the flux distribution $\psi\!(F_{100})$ above the breaks and the distributions of photon index $h\!(\Gamma)$.  The allowed range of the correlation parameter $\beta$ here and in BP1 allows for wider uncertainty in these values in some cases than in MA.  However, as expected, the uncertainty is generally lower here than in BP1, as this sample has roughly twice as many sources.  We note that in this work we find systematically steeper power law slopes of the flux distribution $\psi\!(F_{100})$ below the breaks than in BP1 and MA.  The differences with MA are significant although the differences with BP1 are not significant at the 1$\sigma$ level due to the larger uncertainties.  We also note that the best-fitting widths of the Gaussian fits to the photon index distributions are systematically smaller here than in BP1 and MA for the cases of all blazars and BL Lacs, but not for FSRQs, and that the mean photon index for all blazars found here is higher than in the other works, and this is entirely the result of the mean photon index for FSRQs being higher.

Using the bias free flux distributions we calculated the integrated contribution of FSRQs and BL Lacs to the EGB as a function of flux. We obtain this directly from the cumulative flux distribution which is the main output of the methods used.  Using the simplistic assumption that the distributions continue unchanged to arbitrarily low flux, the best fit contribution of FSRQs and BL Lacs to the total extragalactic gamma-ray radiation in the range from 0.1 to 100 GeV is estimated to be 35\% and 17\% respectively.   The significant uncertainties reported here for the source count slopes and the estimated contribution of blazars to the EGB are ultimately due to the allowed range of the correlation parameter $\beta$, as discussed at length in BP1.  These results for the proportional contribution to the EGB are roughly consistent with a number of other findings.   In \citet{BP2} we consider redshift evolution effects with a smaller sample with available redshifts and find that for FSRQs the contribution to the EGB is 22(+10/-4)\%, while \citet{Marco2} report 30.1(+2.5/-1.7)\% for FSRQs and \citet{Marco3} estimate that BL Lacs contribute between 10\%-15\% of the EGB, with all of these reported here scaled to account for the subsequently reduced estimated level of the EGB as discussed in \S \ref{bgndradsec}.  However, we note that one would not necessarily expect exact agreement between the estimates of total photon output derived here extrapolating the flux distributions to arbitrarily low flux and those derived considering the full redshift evolutions, for two primary reasons.  One is that as the populations clearly have both luminosity and density evolution in redshift \citep[e.g.][]{BP2,Marco2,Marco3}, one would not expect the flux distribution to remain constant at lower fluxes.  The other is that blazars likely do not continue to arbitrarily low flux, indeed in BP1 we argue that a reasonable lower cutoff flux for blazars would be around $\sim10^{-12}$ photons cm$^{-2}$ sec$^{-1}$.  The fact that we do find reasonable agreement between the estimates here and those arrived at considering redshift evolutions indicates not only that analyses seem to be consistent, but also that there may be a ``conspiracy'' where luminosity and density evolution effects, as well as an absolute luminosity cutoff for blazars, and any effects from contributions missed here because some 17\% of 2 LAC blazars are of unclassified type, sum in such a way as to render extrapolations of the distributions recovered here to arbitrarily low fluxes appropriate.  It also points to the utility of comparing this analysis, which is based on the statistically larger 3LAC sample which is the largest extant sample of gamma-ray detected blazars, with those of smaller samples that contain complete redshift information.


\section*{Acknowledgements}

The author acknowledges previous support from NASA-Fermi Guest Investigator grant NNX10AG43G.

\clearpage

\begin{table*}
Table 1 -- Best fitting parameters for {\it Fermi}-LAT blazar intrinsic distributions, as calculated in this work, and two analyses with the smaller {\it Fermi}-LAT 1LAC -- ours \citep[BP1 --][]{BP1} and MA \citep{Marco}
\begin{minipage}{140mm}
\begin{tabular}{lccccccc}

\hline
  & n & $\beta$$^{a}$  & $m_{above}$$^{b}$ & $F_{break}$$^{c}$ & $m_{below}$$^{d}$ & $\mu$$^{e}$ & $\sigma$$^{f}$ \\
\hline
\\
Blazars$^{g}$ (this work) & 774 & -0.13$\pm$0.06 &-2.43$\pm$0.08& 8.0$\pm$0.2 & -1.87$\pm$0.10 & 2.62$\pm$0.05 & 0.17$\pm$0.02 \\
Blazars$^{g}$ (BP1) & 352 & 0.02$\pm$0.08 &-2.37$\pm$0.13 & 7.0$\pm$0.2 & -1.70$\pm$0.26 & 2.41$\pm$0.13 & 0.25$\pm$0.03 \\
Blazars$^{g}$ (MA)        & 352 & -             & -2.48$\pm$0.13 & 7.39$\pm$1.01 & -1.57$\pm$0.09 & 2.37$\pm$0.02 & 0.28$\pm$0.01 \\
\\
BL Lacs (this work)                  & 352 & -0.11$\pm$0.06 & -2.44$\pm$0.09 & 6.0 $\pm$0.2 & -2.00$\pm$0.15 & 2.18$\pm$0.02 & 0.18$\pm$0.03 \\
BL Lacs (BP1)                  & 163 & 0.04$\pm$0.09 & -2.55$\pm$0.17 & 6.5 $\pm$0.5 & -1.61$\pm$0.27 & 2.13$\pm$0.13 & 0.24$\pm$0.02 \\
BL Lacs (MA)                         & 163 & -             & -2.74$\pm$0.30 & 6.77$\pm$1.30             & -1.72$\pm$0.14 & 2.18$\pm$0.02 & 0.23$\pm$0.01 \\
\\
FSRQs (this work)                    & 286 & -0.25$\pm$0.05 & -2.42$\pm$0.09 & 9.2$\pm$0.1 & -1.55$\pm$0.12 & 2.75$\pm$0.06 & 0.19$\pm$0.02 \\
FSRQs (BP1)                    & 161 & -0.11$\pm$0.06 & -2.22$\pm$0.09 & 5.1$\pm$2.0 & -1.62$\pm$0.46 & 2.52$\pm$0.08 & 0.17$\pm$0.02 \\
FSRQs (MA)                           & 161 & -             & -2.41$\pm$0.16 & 6.12$\pm$1.30            & -0.70$\pm$0.30 & 2.48$\pm$0.02 & 0.18$\pm$0.01 \\
\hline

\end{tabular}

$^{a}$The correlation between photon index $\Gamma$ and Log flux $F_{100}$.  See Equation \ref{fdef} and \S \ref{corrs}.  A higher value of $\beta$ (i.e. more positive correlation between flux and photon index absolute value) moves the mean of the photon index distribution down to lower photon index absolute value (lower $\mu$) and makes the faint end source counts slope less steep (less negative $m_{below}$), while a lower value of $\beta$ has the opposite effect.\\
$^{b}$The power law of the intrinsic flux distribution $\psi\!(F_{100})$ at fluxes above the break in the distribution.  See Equation \ref{bpowlaw}.  All values reported for this work include the full range of results and their uncertainties when considering the 1$\sigma$ range of $\beta$.  \\
$^{c}$The flux at which the power law break in $\psi\!(F_{100})$ occurs, in units of $10^{-8}$  photons cm$^{-2}$ sec$^{-1}$.  We present the value even though the precise location of the break is not important for the analysis in this work.  \\
$^{d}$The power law of the intrinsic flux distribution $\psi\!(F_{100})$ at fluxes below the break.  See Equation \ref{bpowlaw}. \\
$^{e}$The mean of the Gaussian fit to the intrinsic photon index distribution $h\!(\Gamma)$. \\
$^{f}$The 1$\sigma$ width of the Gaussian fit to the intrinsic photon index distribution $h\!(\Gamma)$. \\
$^{g}$Including all FSRQs, BL Lacs, and those of unidentified type. \\

\end{minipage}
\end{table*}

\label{lastpage}
\end{document}